\documentclass[aps,pre,reprint,groupedaddress,showpacs]{revtex4-1}

\usepackage{amsmath,amsfonts,amssymb}
\usepackage[dvips]{graphicx}
\usepackage{natbib}
\usepackage[ps2pdf]{hyperref}

\begin{document}

\title{Scaling laws of passive tracer dispersion in the turbulent surface layer.}

\author{Alex Skvortsov}
\email[]{alex.skvortsov@dsto.defence.gov.au}
\affiliation{Defence Science and Technology Organisation,  506
Lorimer Street, Fishermans Bend, Vic 3207, Australia}

\author{Milan Jamriska}
\affiliation{Defence Science and Technology Organisation,  506
Lorimer Street, Fishermans Bend, Vic 3207, Australia}

\author{Timothy C. DuBois}
\affiliation{Defence Science and Technology Organisation,  506
Lorimer Street, Fishermans Bend, Vic 3207, Australia}

\date{\today}

\begin{abstract}

Experimental results for passive tracer dispersion in the turbulent
surface layer under stable conditions are presented. In this case, the
dispersion of tracer particles is determined by the interplay of
three mechanisms: relative dispersion (celebrated Richardson's
mechanism), shear dispersion (particle separation due to
variation of the mean velocity field) and specific surface-layer
dispersion (induced by the gradient of the energy dissipation rate
in the turbulent surface layer). The latter mechanism results in the
rather slow (ballistic) law for the mean squared particle
separation. Based on a simplified Langevin equation for particle
separation we found that the ballistic regime always dominates at
large times. This conclusion is supported by our extensive
atmospheric observations. Exit-time statistics are derived from the
experimental dataset and show a reasonable match with the simple
dimensional asymptotes for different mechanisms of tracer
dispersion, as well as predictions of the multifractal model and
experimental data from other sources.

\end{abstract}

\pacs{47.27.nb, 47.27.eb, 64.60.al, 05.45.Tp, 92.60.Fm, 92.60.Mt}

\maketitle

\section{Introduction}

The phenomenon of scalar turbulence or random advection of tracer
particles by random turbulent flow has become a topic of significant
attention during the last few years \cite{Frisch_1995},
\cite{Shraiman_2000}, \cite{Falkovich_2001}, \cite{Duplat_08},
\cite{Celani_04}, \cite{Celani_05}, \cite{Gioia_04}. This is not
only due to its significance for understanding the transport
processes in global geophysical systems (e.g.\ the atmosphere and
oceans) and as a theoretical framework for the design of some
technological applications (mixers, chemical reactors, combustion
chambers), but is perhaps even more important as an instructive example
of modern methods of theoretical physics applied to a highly
non-equilibrium dynamic system in order to deduce new phenomenology
and a wealth of analytical results that can be validated numerically
and experimentally. Known examples are: analytical scaling for
white-noise scalar turbulence (solutions of the Kraichnan model
\cite{Frisch_1995}, \cite{Shraiman_2000}, \cite{Falkovich_2001});
application of conformal invariance to the two-dimensional tracer
flow \cite{Duplat_08}; re-normalization group formalism;
multi-fractal structure of tracer statistics
\cite{Frisch_1995} and others (see review \cite{Boffetta_08} and
references therein). Remarkably, the recent applications of the
theoretical framework for this phenomena provide rigorous ways to
overcome the initially restrictive assumptions of the underlying
model of locally isotropic turbulence by incorporating effects of
anisotropy, flow boundaries, mean velocity shear, buoyancy, etc.\ and
enable the analytical calculations of associated corrections
\cite{Lebedev_04}, \cite{Celani_04}, \cite{Bistagnino_08}.

In this paper we  report our experimental results on the
dispersion of passive tracers in the turbulent surface layer. The
wall-bounded turbulent flow provides flexible settings to study the
effects of mean velocity shear (i.e.\ non-uniform wind) and system
boundaries (the underlying surface) on the passive scalar
dispersion.

The celebrated Richardson law \cite{Richardson_26} established the
growth of the mean inter-particle distance:
\begin{equation}
 \label{eqn:E:01}
\langle{R^{2}(t)}\rangle \sim  \lambda t^{n},
\end{equation}
where $n=3$, $\lambda$ is a scale-independent dimensional parameter,
and became a signature of turbulent dispersion (see review
\cite{Salazar_2009}). It was later recognized by Oboukhov and Corsin
that this law is a direct consequence of Kolmogorov scaling in
turbulence \cite{Monin_1975}, \cite{Falkovich_2001}. Indeed, from a
power-law assumption for the velocity differences  $\delta v \sim
d{R}/dt \propto R^{h}$ and (\ref{eqn:E:01}) it is easily to derive
$R \propto t^{\frac{1}{1-h}} $ or $h = 1 - 2/n$. The latter
expression leads to the Richardson law with
$\langle{R^{2}(t)}\rangle \propto
 t^{3}$ for the Kolmogorov scaling $h=1/3$. This is a
reflection of an intimate and well-known connection between the
power exponent of the particle separation law $n$ in
(\ref{eqn:E:01}) and the passive tracer statistics
\cite{Falkovich_2001}. From a mathematical point of view, this
connection can be translated into a scaling law of the correlation
functions of the tracer concentration that includes the parameter
$n$ \cite{Monin_1975}, \cite{Falkovich_2001}:
\begin{eqnarray}
\label{e:assimp_eq}
 S_2(R) = 2[F_2(0) - F_2(R)] \propto R^{2/n}, ~~ n = 1/(1-h),
\end{eqnarray}
where
\begin{eqnarray}
\label{e:assimp_eq1} F_2(R) =  \langle C (\textbf{x} + \textbf{R}, t)
C (\textbf{x}, t)\rangle
\end{eqnarray}
is the pair-correlation function.

Different values of $n$ (the scaling exponent of the mean squared
displacement) have been derived theoretically and experimentally for
different kinds of turbulent flow  (see \cite{Salazar_2009},
\cite{Celani_05} and refs therein). Contrarily, a particular value
of $n$ can be associated with a particular energy injection
mechanism of turbulence and can be used for characterization of its
dispersive properties. As was mentioned above, for the tracer
dispersion by Kolmogorov (locally isotropic) turbulence, $n=3$
(Richardson law). For Bolgiano-Obukhov (buoyancy dominated)
turbulence, $n=5$ \cite{Bistagnino_08}. For turbulence with mean
velocity shear it was recently deduced \cite{Celani_05} that $n=6$
(for particle separation along the mean velocity) and $n=4$ (for
separation in the transverse direction). It is worth noting that
in all latter cases the separation of particles is always faster
than in the standard Richardson regime (i.e.\ without velocity
shear). For surface layer turbulence (i.e.\ turbulent motion
near a flow boundary), whose scaling properties are significantly
different from the isotropic Kolmogorov model,  the value of $n$ can
eventually deviate from Richardson's prediction. Particularly, the
self-similarity arguments applying to this case  lead to  much lower
values of the scaling exponent of the mean squared displacement
($n=2$ or so-called ``ballistic'' regime, since $R \propto t$)
\cite{Mikkelsen_02}. Finally, for the ``combined'' case (turbulent
flow near a boundary \emph{and} a mean velocity shear) Ref
\cite{Lebedev_04} argues that there is no universal value for the
parameter $h$ in (\ref{e:assimp_eq}) (and hence neither for exponent
$n$).

From simple dimensional analysis it is obvious that the dimensional
parameter $\lambda$ in (\ref{eqn:E:01}) should also be different for
the different values of $n$, i.e.\ for the particular type of
turbulent flow.  For instance, in the case of Kolmogorov turbulence
$\lambda = \epsilon$ ($\epsilon$ is the kinetic energy flux); but
for surface layer turbulence, the friction velocity $v_*$ (and
not the dissipation rate $\epsilon$) becomes the only similarity
parameter \cite{Monin_1975}, $\lambda = {v_*}^2$ (therefore the
expression $\lambda t^2 = (v_*t)^2$ provides the dimension of
$R^2$).

The same dimensional analysis also  leads to the important
conclusion that the ballistic regime (and not the shear effect) is
the universal long-time asymptote of the dispersion process of the
tracer particle in the turbulent surface layer (see
\cite{Mikkelsen_02} and refs therein). This can be easily seen based
on the following arguments. Let us assume that in addition to $v_*$,
the problem is characterized by a length scale $l$ (i.e.\ initial
separation of the particles or the initial distance to the ground).
Then the dimensional arguments translate (\ref{eqn:E:01}) into
\begin{equation}
 \label{eqn:E:051}
\langle{R^{2}(t)}\rangle \sim  \lambda t^{n}, ~~ \lambda = v_*^n
l^{2 - n}.
\end{equation}
A na\"ive conjecture at this point would be that a dispersion
mechanism with the highest value of $n$ (e.g.\ shear)  will dominate
over large time periods (i.e.\ at $t \geq l/v_*$). However, more
careful analysis casts doubt on this conclusion. Indeed, for $t \gg
l/v_*$ particles should ``forget'' about their initial positions and
the parameter $l$ should be dropped from the expression for the
particle separation (\ref{eqn:E:051}) all together. We can see that
the only possibility for this is the case when $n=2$: conversion  to
the ballistic regime.

This variety of possible scenarios of tracer dispersion in the
surface layer turbulence and some ambiguity of available analytical
predictions (at least when being straightforwardly applied)
motivated our experimental study of this phenomenon.

\section{Experimental Procedure}

Scrutinizing measurements at a given location within the Taylor
hypothesis of frozen turbulence $R = Ut$, where $U = const$; it
follows from (\ref{eqn:E:01}) that:
\begin{equation}
\label{eqn:E:04} S_2(R) = 2[F_2(0) - F_2(R)] \propto R^{2/n} \propto
(Ut)^{2/n},
\end{equation}
so single point time-measurements of concentration also show
power-law behavior with the same characteristic exponents as the
spatial measurements.

We estimated the value of parameter $n$ in (\ref{eqn:E:04}) based on
extended atmospheric observations of the tracer concentrations. A
continuous (and relatively stable) influx of particles into the
turbulent flow was supplied by the anthropogenic activity in the
surrounding urban areas (about 20 km in size). This ``highly
distributed'' source of particles maintained the quasi-equilibrium
(and on average, well mixed) tracer distribution which was validated
by our long-term observations. The observation tower was at a height
of $H$ = 12m and the tracer concentration was measured by means of
scattered light intensity off incoming particles. Continuous
observation was undertaken over a period of 21 days with an air
sample taken every 5 seconds (30 liters per sample). The
instrumentation provided total particle count as well as their size
distribution in the range of 1 - 10 $\mu m$. Local meteorological
observations (three conventional meteorological stations in the
vicinity) and global observations (with resolution about 1 km) were
available to draw the conclusions about local meteorological
conditions of the surface layer (stability, profiles of wind speed
and temperature). Wind speed and temperature were also measured
directly at the sampling point ($H$ = 12m).

As usual in meteorological studies, we anticipated that the night
time observations would correspond to the case of the boundary layer
under stable or neutral conditions. For convective turbulence to be
generated, an intensive heat flux from the underlying surface has to
occur and this, of course, should include the significant effect of
solar radiation. Indeed, in our experimental results the latter
condition is bound to the daytime observations.

\section{Data Analysis}

\begin{figure}[h]
    \includegraphics[width=\columnwidth]{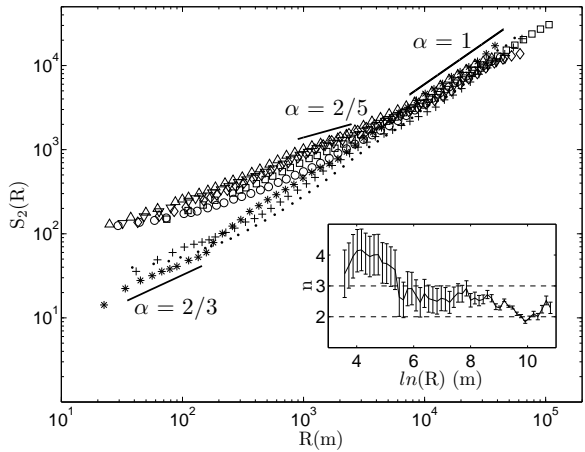}
    \caption{Structure function of the tracer concentration $S_2(R)  \propto R^{\alpha} \propto U_{H}t^{\alpha}
    $ on a log-log scale.
Inset shows the estimation of how the scaling exponent of the mean
inter particle displacement $\langle{R^{2}(t)}\rangle \propto t^{n}$
converges over time. Each data set corresponds to a maximum of $4$ hours of
observations and wind velocities outlined in Table \ref{table:01}: 
$+$, Sample $1$; $\circ$, Sample $2$; $\ast$, Sample $3$; $\cdot$, Sample $4$; 
$\diamond$, Sample $5$; $\square$, Sample $6$; $\lozenge$, Sample $7$; $\vartriangle$, Sample $8$ and $\triangledown$, Sample $9$.
$\alpha = 2/3$ is the Richardson regime, $\alpha
\approx 2/5$ is the velocity shear mechanism and $\alpha = 1$ is the
ballistic regime. Global convergence to the ballistic regime
($\alpha =1$) is clearly visible.}
    \label{fig:01}
\end{figure}

Undertaking correlation analysis of the time series of the
concentration $C(t)$, of single point measurements, we found that
the minimum time span that provides reliable statistics of the
process should be more than three hours. As the first step, we
plotted $S_2(R)$ as a function of $R = Ut$ on a log-log scale and
evaluated the scaling exponent $n$ from the expression  $S_2(R)
\propto (Ut)^{2/n}$ predicted by (\ref{e:assimp_eq1}) and
(\ref{eqn:E:04}). Some examples of these plots are depicted in Fig
\ref{fig:01}.

In this paper we report on the analysis of the data that corresponds
to the stable conditions of the turbulent surface layer
(results on convective tracer dispersion will be published
elsewhere). As was mentioned above, the turbulent motion in this
case is determined only by a one dimensional parameter - the
friction velocity $v_*$, which provides the scale for velocity
fluctuations. This self-similarity property of the surface layer
turbulence is expressed by the well-known relations
\cite{Monin_1975}
\begin{equation}
\label{eqn:E:05} \frac{\text{d} U(z)}{\text{d} z } =
\frac{v_*}{\kappa z}, ~~ \epsilon = \frac{v^3_*}{\kappa z}, 
\end{equation}
where $z$ is the vertical distance from the underlying surface and
$\epsilon$ is the dissipation rate, $\kappa =0.4$.

The values of wind speed relevant to the data series depicted in Fig. \ref{fig:01} are listed in Table \ref{table:01}. By using the
logarithmic profile $U(z)=(v_*/\kappa)ln(z/z_0)$ followed from (\ref{eqn:E:05}) we can estimate $v_* = c
U_H$ ($c = {\kappa}/ \ln ({H}/{z_0})$) and then $\epsilon  =
{v^3_*}/{\kappa z_0}$. Assuming a typical value of roughness height $z_0 \approx 0.1$ m leads to
to the estimates $c \approx 0.01$ and $\epsilon \approx 0.004$ kg m\textsuperscript{2}/s\textsuperscript{3} for $U_H = 5.52$ m/s.
The parameter $T_H \sim H/U(H) \sim  H/cv_*$ corresponds to the time
scale for a particle to reach the underlying surface and was of
order $10^2$ seconds.

\begin{table}[h]
     \caption{\label{table:01} Apropos mean wind speeds $U_H$ for Fig. \ref{fig:01} }
\begin{ruledtabular}
\begin{tabular}{ccrc}
& Sample & $U_H$ (m/s) & \\
\hline
& 1. & $4.15\pm 0.12$ & \\
& 2. & $2.94\pm 0.11$ & \\
& 3. & $2.25\pm 0.30$ & \\
& 4. & $3.85\pm 0.12$ & \\
& 5. & $5.52\pm 0.16$ & \\
& 6. & $7.42\pm 0.05$ & \\
& 7. & $4.27\pm 0.21$ & \\
& 8. & $2.49\pm 0.06$ & \\
& 9. & $3.20\pm 0.12$ & \\
\end{tabular}
\end{ruledtabular}
\end{table}

The low relative fluctuations of the wind speed was one of the reasons
that these data series were selected for analysis. It was also a justification
for the application of the Taylor hypothesis to these datasets.
It is worth noting that a particular value of the mean wind velocity has no relevance to the main
reported result of our paper; it will only change the point at which
a data series will approach the global asymptote in Fig. \ref{fig:01}. It
will not change either the existence of, nor the slope of this asymptote.

The dynamics of particle separation in the turbulence surface layer
can be described by modifying the Langevin equation for particle
separation \cite{Sawford_06}. By including the shear term proposed
in \cite{Celani_05} we can arrive at the following system:
\begin{eqnarray}
 \label{eqn:E:052}
 {d{R}_x} = G(R_z) dt +  D(R) dt  +  \sqrt{2K} ~dW, \\
 {d {R}_\perp} =  D(R)dt  +  \sqrt{2K} ~dW,
\end{eqnarray}
where $G(R_z) = (dU /dz) R_z$ is the velocity shear term, $D(R) = (
dK/dR + 2K/R)$ is the drift term, $K(R) = \beta \epsilon^{1/3}
R^{4/3}$ is the diffusivity, $\beta = const$, $\epsilon(\textbf{R})$
is the local dissipation flux (which describes the local activity of
the turbulent flow) (\ref{eqn:E:05}), $dW$ is an isotropic Brownian
motion, $R_z$ is the vertical component of separation, $R_x$ is its
horizontal component and ${\textbf{R}}_\perp = [R_z, R_y]$ are the
components perpendicular to the mean shear; so $R^2 = {R}^{2}_x +
\textbf{{R}}^{2}_\perp $.

For $t\ll T_H$, $U'(z) \sim v_*/\kappa H$,
$\epsilon(\textbf{R})\approx \epsilon(H) = const$, so $G \ll D$. By
placing  $R_z \sim R_x \sim R_\perp \sim R$, we recover the standard
Richardson regime $R^2 \propto t^3$ \cite{Celani_05}. For the
intermediate times ($t \approx T_H$) the velocity shear mechanism
dominates ($G \gg D$) and following arguments of \cite{Celani_05} we
arrive at faster separation of particles ($R^2 \propto t^n$ with $ 4
\leq n \leq 6$). Finally, at the limiting case $t\geq T_H$ the
separation becomes slower (ballistic with $R^2 \propto t^2$). The
latter can be deduced from the system (\ref{eqn:E:052}). At $t\gg
T_H$ we can use estimates: $U'(z)\sim v_*/\kappa R_z$,
$\epsilon(\textbf{R})= {v^3_*}/{\kappa R_z}$, which leads to a
simplified system for the long-time asymptotes
\begin{eqnarray}
 \label{eqn:E:053}
{d{R}_x}  \propto \sqrt{v_* R}~ dW  , ~~ {d {R}_\perp} \propto
\sqrt{v_* R}~ dW,
\end{eqnarray}
since $R_z \sim {R}_\perp \sim R$.

We can see from the diffusive equation that the associated pdf
$p(R,t)$ can be derived
\begin{eqnarray}
 \label{eqn:E:0531}
\frac{\partial p}{\partial t} = \gamma v_* \frac{\partial }{\partial
R} \left(R \frac{\partial p }{\partial R} \right),
\end{eqnarray}
where $\gamma$ is the constant of order of unity. This equation has
a straightforward solution:
\begin{eqnarray}
 \label{eqn:E:0532}
p(R,t) = \frac{1}{2 \gamma v_* t} \exp \left ( - \frac{R} {\gamma
v_* t} \right ).
\end{eqnarray}
It is evident from here that the main asymptote of the model as $t
\rightarrow \infty$ is the ballistic regime  $\langle R (t) \rangle
= \int^{\infty}_0 R p(R,t) \text{d}R  \sim v_* t$.

The existence of this  global ballistic asymptote often suppresses
the effect of the shear dispersion and clearly emerges from our
experimental data (see Fig 1). We observe that at the short-time
intervals, close to $t = 0$, the parameter $n \approx 3$ (Richardson
regime), then it reaches maximum: $n \approx 5$ (shear dispersion)
and finally $n$ gradually decays to the the ballistic value: $n
\approx 2$. It is worth noting that the long-time ballistic limit $n
\approx 2$ is in agreement with recent experimental data on
atmospheric dispersion $2 \leq  n < 3$, see \cite{Mikkelsen_02},
\cite{Salazar_2009} and references therein.

The non-trivial scaling properties of tracer dispersion in the turbulent
surface layer can be also demonstrated by means of
exit-time statistics for the concentration time series
\cite{Boffetta_08}; which are statistics of time intervals in
which a measured value of concentration exits through a set of
thresholds. By scanning the time series for a given threshold
$\delta C$, one can recover a set of times $\tau_i(\delta C)$, where
the measured concentration reaches this threshold. This set can then
be used to calculate the Inverse Structure Function (for details see
\cite{Boffetta_08}, \cite{Bistagnino_08}):
\begin{equation}
\label{eqn:E:06} \Sigma_q (\delta C) \equiv  \langle \tau^q(\delta
C) \rangle.
\end{equation}

A comprehensive analysis of the properties of the Inverse Structure
Function can be fulfilled by applying the well-known mutifractal
approach \cite{Boffetta_08}, \cite{Schmitt_05}. This results in the
following scaling \cite{Boffetta_08}, \cite{Schmitt_05}:
\begin{equation}
\label{eqn:E:063} \Sigma_q (\delta C) \propto (\delta
C)^{\chi(q)},~~ \chi(q) = \min_h [{({q + 3 - D(h)})}/{h}]
\end{equation}
where $h$ is the index of singularity from the range $[h_{min},
h_{max}]$ such that $t \sim (\delta C)^{h}$ and $D(h)$ is the fractal
dimension of the set with a singularity $h$. It is worth noting that
particular values $h=1/3$, $h=1/5$ and $h=1/2$ correspond to the
Richardson law, the  shear dispersion and the ballistic scaling
discussed above.

According to \cite{Schmitt_05}, the function $(q + 3 - D(h))/h$
reaches its minimum at the upper boundary of the singularities'
range, so we can set $h=h_{max}$ and write
\begin{eqnarray}
\label{eqn:E:064}
 \chi (q) = (q + 3 - D(h_{max}))/h_{max}.
\end{eqnarray}

\begin{figure}[h]
    \includegraphics[width=\columnwidth]{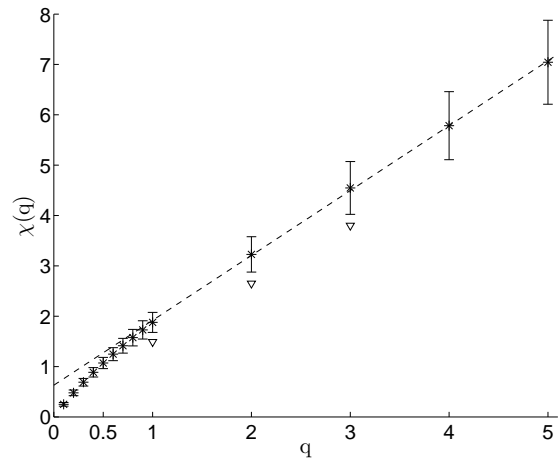}
    \caption{Mean Inverse Structure Function for 12 experimental data sets ($*$).
    Error bars correspond to $\pm$ mean standard deviation. The dashed line
    represents a linear best fit over $1 \le q \le 5$ predicted by (\ref{eqn:E:064}) \cite{Schmitt_05}. Experimental values
    for temperature fluctuations ($\triangledown$) of \cite{Beaulac_04} are presented for
    reference.}
    \label{fig:02}
\end{figure}

This leads to the important conclusion that $ \chi (q)$ is a linear
function of $q$ (i.e.\ no intermittency correction), which has been
verified by numerical simulations and by some limited experimental
data \cite{Schmitt_05}. This conclusion was also validated against
our experimental dataset and the results are shown in Fig
\ref{fig:02}. We observe that indeed $ \chi (q)$ closely follows
the predicted linear trend for $q \geq 1$ and provides a reasonable
match with the experimental data available in the literature
\cite{Beaulac_04}. A change in slope near the value $q=1$ can be
attributed to a contribution of the slow (differentiable) components
of turbulent motion \cite{Bistagnino_08}. A model for this effect
will be discussed elsewhere.

\begin{figure}[h]
    \includegraphics[width=\columnwidth]{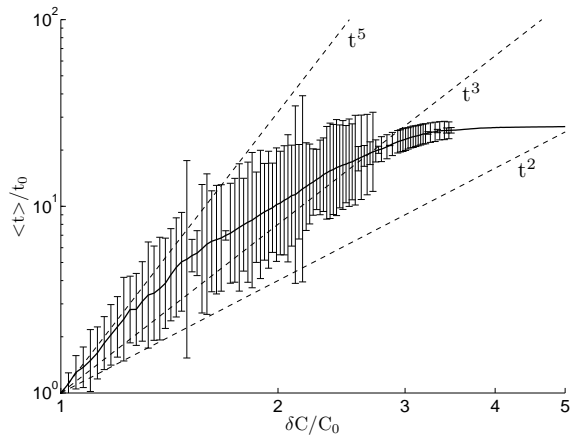}
    \caption{Exit-time for the first moment of the concentration time series.
    Different regimes of dispersion correspond to the different power-law asymptotes (see text).
    Scaling factors $C_{0}$, $t_{0}$ were used for collapsing data thus a better visual appearance (see text).}
    \label{fig:03}
\end{figure}

We also present a plot of  the mean exit-time $\Sigma_1 (\delta C) =
\langle t(\delta C) \rangle$ as a function of the concentration
threshold $\delta C$ in Fig \ref{fig:03}. $C_{0}$ and its associated
$t_{0}$ are located at the concentration minima for each sample, and
are used to collapse data only. Three asymptotes corresponding to
the regimes discussed above are also depicted for comparison. These
asymptotes can be easily established from (\ref{eqn:E:04}) based on
simple dimensional arguments. Indeed, for the short times $t \ll
T_H$ we can assume Richardson (or Corsin-Obukhov) scaling
$\delta C \propto t^{1/3}$ and derive $\langle t(\delta C) \rangle
\propto (\delta C)^3$. Similarly, for the shear dispersion we can
write $\langle t(\delta C) \rangle \propto (\delta C)^n, n \approx
5$ \cite{Celani_05}; while for the ballistic regime at the longer
times ($t \gg T_H$) we arrive at the scaling $\langle t(\delta C)
\rangle \propto (\delta C)^2$. A subsequent change of regime as
discussed above, corresponding to the curve  $\langle t(\delta C)
\rangle$ deviating from one asymptote to another, is clearly
visible. From this plot it is also  evident why in the case of
wall-bounded turbulent dispersion it is not possible to assign any
universal value for parameter $n$ in the scaling law $\langle
t(\delta C) \rangle \propto (\delta C)^n$ (similarly in
(\ref{e:assimp_eq})), which is in agreement with the conclusions of
\cite{Lebedev_04}.

\section{Conclusions}

We presented experimental results for passive tracer dispersion in
the turbulent surface layer under stable conditions. In this case the
dispersion of tracer particles was affected by the mean velocity
gradient and flow boundaries. We found that our observations can be
intrinsically explained with the three-stage model of tracer
dispersion. During the first stage of separation ($T \ll H/v_*$,
where $H$ is the observation height and $v_*$ is the friction
velocity) tracer particles obey the standard Richardson model.
During the second stage ($T \approx H/v_*$) the shear mechanism of
dispersion dominates \cite{Celani_05}. Finally, when ($T \gg H/v_*$)
the shear mechanism is followed by a transition to the ballistic
regime of dispersion induced by the specific scaling properties of
the turbulence in the surface layer. This scenario of inter-particle
distance seems to be in agreement with atmospheric observations
\cite{Mikkelsen_02}, \cite{Salazar_2009} as well as experimental
results on tracer dispersion by turbulent surface flow in water
channels \cite{Borgas_10}. We found that the ballistic regime
always suppressed the velocity-shear effect predicted in
\cite{Celani_05} at the later stage of dispersion, resulting in much
slower rate of particle separation (i.e.\ lower value of parameter
$n$ in (\ref{eqn:E:01})). Exit-time statistics were derived from the
experimental dataset and showed a reasonable match with the simple
dimensional asymptotes, as well as predictions of the multifractal
model and experimental data from other sources.

\bibliography{particles}

\end{document}